\input epsf.tex    

\documentstyle[12pt]{article}

\newcommand{\bmat}{\left(\begin{array}}
\newcommand{\emat}{\end{array}\right)}

\def\yzero{\smash{\hbox{$y\kern-4pt\raise1pt\hbox{${}^\circ$}$}}}

\def\beq{\begin{equation}}
\def\eeq{\end{equation}}
\def\beqa{\begin{eqnarray}}
\def\eeqa{\end{eqnarray}}

\def\-{\hphantom{-}}
\def\ov{\overline}
\def\s2{\frac{1}{2}}

\def\beq{\begin{equation}}
\def\eeq{\end{equation}}
\def\beqa{\begin{eqnarray}}
\def\eeqa{\end{eqnarray}}
\def\tr{{\rm tr \,}}

\def\IF{\relax{\rm I\kern-.18em F}}
\def\II{\relax{\rm I\kern-.18em I}}

\def\cp{{\cal P}}
\def\IC{\bf C}
\def\IZ{\bf Z}

\def\IP{\bf P}
\def\IT{\bf T}
\def\IM{\bf M}
\def\IX{\bf X}
\def\z2z2{$\IC^3/(\IZ_2\times\IZ_2)$}

\def\NN{{\cal N}}
\def\Dsl{\,\raise.15ex\hbox{/}\mkern-13.5mu D} 

 \def\cp#1{\relax\ifmmode {\IP\kern-2pt{}_{#1}}\else $\IP\kern-2pt{}_{#1}$\=fi}
\newcommand{\drawsquare}[2]{\hbox{%
\rule{#2pt}{#1pt}\hskip-#2pt
\rule{#1pt}{#2pt}\hskip-#1pt
\rule[#1pt]{#1pt}{#2pt}}\rule[#1pt]{#2pt}{#2pt}\hskip-#2pt
\rule{#2pt}{#1pt}}

\newcommand{\Ysymm}{\raisebox{-.5pt}{\drawsquare{6.5}{0.4}}\hskip-0.4pt%
        \raisebox{-.5pt}{\drawsquare{6.5}{0.4}}}
\newcommand{\Yasymm}{\raisebox{-3.5pt}{\drawsquare{6.5}{0.4}}\hskip-6.9pt%
        \raisebox{3pt}{\drawsquare{6.5}{0.4}}}

\begin{document}

\pagestyle{empty}
\vspace*{.5in}
\rightline{IFT-UAM-CSIC-02-03}
\rightline{\tt hep-th/0301032}
\vspace{2cm}
 
\begin{center}
\LARGE{\bf Chiral four-dimensional string compactifications with 
intersecting D-branes \\[10mm]}
\medskip
\large{Angel M. Uranga \footnote{\tt Angel.Uranga@uam.es} \\[2mm]}
I.M.A.F.F. and Instituto de F\'{\i}sica Te\'orica C-XVI \\
Universidad Aut\'onoma de Madrid, 28049 Madrid, Spain \\ [3mm]
\end{center}
 
\smallskip

\begin{center}
\begin{minipage}[h]{14.5cm}
{\small
We review the construction of chiral four-dimensional compactifications of 
type IIA string theory with intersecting D6-branes. Such models lead to 
four-dimensional theories with non-abelian gauge interactions and charged 
chiral fermions. We discuss the application of these techniques to 
building of models with spectrum as close as possible to the Standard 
Model, and review their main phenomenological properties. We also 
emphasize the advantages/disadvantages of carrying out this idea
using supersymmetric of non-supersymmetric models.}
\end{minipage}
\end{center}

\newpage                                                        

\setcounter{page}{1} \pagestyle{plain}
\renewcommand{\thefootnote}{\arabic{footnote}}
\setcounter{footnote}{0}
 
\section{Introduction}      

String theory has the remarkable property that it provides a description 
of gauge and gravitational interactions in a unified framework 
consistently at the quantum level. It is this general feature (beyond 
other beautiful properties of {\em particular} string models) that makes 
this theory interesting as a possible candidate to unify our description 
of the different particles and interactions in Nature.

Now if string theory is indeed realized in Nature, it should be able to 
lead not just to `gauge interactions' in general, but rather to gauge 
sectors as rich and intricate as the gauge theory we know as the Standard 
Model of Particle Physics. This is described by a gauge group
\beqa
SU(3)_c\times SU(2)_w \times U(1)_Y
\eeqa
a set of charged chiral (left-handed) fermions in three copies with 
identical gauge quantum numbers, namely
\beqa
3 & \times & [\, (3,2)_{1/6}\, +\, ({\ov 3},1)_{1/3}\, + \, ({\ov 
3},1)_{-2/3} \, + \nonumber \\
&& +\, (1,2)_{-1/2} \, + \, (1,1)_{1}\, + \, (1,1)_0 \, ]
\eeqa
(where we have also included the right-handed neutrinos), and a scalar 
particle, the Higgs multiplet, with gauge quantum numbers
\beqa
(1,2)_{-1/2}
\eeqa
These particles have also very specific interactions. They correspond to 
all possible terms consistent with the gauge symmetries, and with some 
`accidental' global symmetries, like baryon or lepton number. 

\medskip

This quantum field theory has certain characteristic and very interesting 
rough features, for instance the very existence of non-abelian gauge 
interactions, the presence of charged chiral fermions, the spontaneous 
breaking of gauge symmetries, the replication of fermion families, etc.
On the other hand, once this basic structure is fixed, the model contains
a large number of more detailed features, for instance the particular 
values of gauge couplings or the pattern of yukawa interaction strengths, 
etc.

All these features are simply unexplained by the Standard Model itself, 
where they are external inputs. It is a natural question to ask whether it 
is possible to explain or reproduce them in a microscopic underlying 
theory, like string theory. In a sense to be qualified below, this is 
the purpose of the branch of string theory known as String Phenomenology.

Clearly, even if string theory is correct, it is not realistic to hope to 
construct explicitly `the' string theory of the world, in the following 
sense. As is well known, the simplest string theories propagate in 
spacetimes of ten dimensions (eleven for M-theory), and usually have a 
high degree of supersymmetry. In the process of describing models reducing 
at low energies to four dimensions and low or no supersymmetry, there is 
an enormous arbitrariness in the choice of the background configuration, 
namely the compactification data, etc. We do not have a good enough 
understanding of whether any of these is in any sense preferred over the 
rest (either dynamical, cosmological or anthropically). Moreover, most of 
the regimes of string theory are not accessible to our computational tools 
(which involve perturbation theory in spacetime string coupling, and in 
$\alpha'$ corrections around exactly solvable conformal field theory 
backgrounds). These considerations imply that it is extremely unlikely that 
we find the `correct' string theory by trial and error.

From this viewpoint, the purpose of String Phenomenology (for this talk) 
is more modest, but still non-trivial and hopefully achievable:

{\bf I.} We should describe (and hopefully classify) different {\em 
setups} in string theory, which lead to the same kind of physics as the 
Standard Model (at the rough level, namely leading to non-abelian gauge 
interactions, replicated chiral fermions, etc).

{\bf II.} Within each setup, we should construct explicit examples with 
low energy physics as close as possible to the Standard Model. From these 
particular examples we should extract generic, robust, features of their 
low-energy phenomenology, which can therefore be considered as natural 
predictions of the setup.

Once this program is fulfilled to a satisfactory level (and we are still 
far from understanding diverse issues, e.g. related to the absence of 
exact supersymmetry at low-energy), future experimental data should be 
able to start cornering what kind of string theory may be underlying our 
world.

\medskip

{\bf Prototypical example: Compactification of heterotic string}

At this point it will be useful to provide the most notorious example of 
what we mean by a `setup', and what lessons can be drawn from it.
The most familiar string theory setup reproducing the rough features 
of the Standard Model at low energies is compactification of the heterotic 
string theories (although we center on perturbative heterotic models, 
similar features are also valid for compactifications of Horava-Witten 
theory, or in non-perturbative heterotic vacua with five-branes).

\smallskip

{\bf I. Construction}

The heterotic string leads to a highly supersymmetric massless sector 
containing gravitational and gauge interactions (with gauge group $G$ 
being $E_8\times E_8$ or $SO(32)$), propagating in ten-dimensional 
spacetime. 

To obtain four-dimensional physics, and to reduce the amount of 
supersymmetry, we choose the background spacetime to be $\IM_4\times 
\IX_6$, with $\IX_6$ a compact six-dimensional manifold (usually chosen 
Calabi-Yau to preserve 4d $\NN=1$ supersymmetry). Moreover, to reduce the 
4d gauge group, we turn on a non-trivial background for the internal 
components of gauge fields in a subgroup $H\subset G$). The 4d gauge group 
is given by the commutant of $H$ in $G$, namely the elements of $G$ 
commuting with $H$ (for instance, for $G=E_8\times E_8$, the choice 
$H=SU(3)$ leads to a 4d gauge group $E_6\times E_8$; while $H=SU(3)\times 
\IZ_2$ can lead to a 4d gauge group $SU(3)\times SU(2)\times U(1)$).
Fig. \ref{heterotic} shows configurations of this kind.

\begin{figure}
\begin{center}
\centering
\epsfysize=3.5cm
\leavevmode
\epsfbox{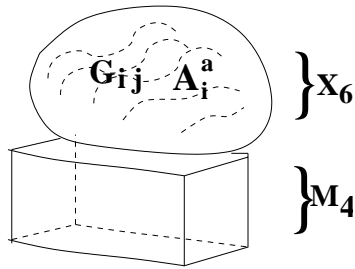}
\end{center}
\caption[]{\small Picture of heterotic string compactification.}
\label{heteroticworld}
\end{figure}     

Finally, these backgrounds ensure that the low-energy theory naturally 
contains 4d chiral fermions charged under the gauge group, replicated a 
number of times determined by the topology of the compactification space 
$\IX_6$ and of the gauge background (for instance, for the choice 
of internal gauge background known as standard embedding, with $H=SU(3)$, 
the number of families, i.e. 27's of $E_6$ is given by the Euler 
characteristic $\chi(\IX_6)$ of $\IX_6$).

\medskip

{\bf II. Generic features/natural predictions of the setup}

Explicit heterotic models of this kind, with spectrum very close to that 
of the Standard Model, have been constructed using different techniques 
(Calabi-Yau compactifications with non-standard embedding, orbifolds, 
fermionic constructions, etc). This allows to extract certain generic 
features of this setup.

$\bullet$ A very important one is the value of the string scale in this 
setup, which follows form analyzing the strength of gravitational and 
gauge interactions, as we quickly review.

The 10d gravitational and gauge interactions have the structure
\beqa
\int\, d^{10}x \, \frac{M_s^8}{g_s^2}\, R_{10d} \quad ;\quad
\int\, d^{10}x \, \frac{M_s^6}{g_s^2}\, F_{10d}^{\; 2} 
\eeqa
where $M_s$, $g_s$ are the string scale and coupling constant, and 
$R_{10d}$, $F_{10d}$ are the 10d Einstein and Yang-Mills terms. Upon 
Kaluza-Klein compactification on $\IX_6$, these interactions reduce to 
4d and pick up a factor of the volume $V_6$ of $\IX_6$ 
\beqa
\int\, d^{4}x \, \frac{M_s^8\, V_6}{g_s^2}\, R_{10d} \quad ;\quad
\int\, d^{4}x \, \frac{M_s^6 \, V_6}{g_s^2}\, F_{10d}^{\; 2} 
\eeqa
From this we may express the experimental 4d Planck scale and gauge 
coupling in terms of the microscopic parameters of the 
string theory configuration
\beqa
M_P^2 \, = \, \frac{M_s^8\, V_6}{g_s^2} \, \simeq \, 10^{19}\, {\rm GeV} 
\quad ; \quad
\frac{1}{g_{YM}^2} \, = \, \frac{M_s^6\, V_6}{g_s^2} \, \simeq \, {\cal 
O}(.1)
\eeqa
From these we obtain the relation
\beqa
M_s \, = \, g_{YM} \, M_P \, \simeq\, 10^{18} \, {\rm GeV}
\eeqa
which implies that the string scale is necessarily very large in this kind 
of constructions.

$\bullet$ In order to protect the hierarchy between this fundamental scale 
and the weak scale $M_w\simeq 10^2$ GeV, the requirement of $\NN=1$ susy 
at intermediate scales is quite essential.

$\bullet$ The large fundamental scale helps in protecting the proton 
against too fast decay, by suppressing higher dimension operators 
violating baryon number.

$\bullet$ Gauge coupling constants unify at the string scale, and lead to 
values in relatively good agreement with measured low-energy 
couplings.

These and many other interesting features make this setup an extremely 
appealing picture of our world. It is important however to mention that 
our understanding of it is not fully satisfactory, and that difficult 
open questions remain, like the breaking of supersymmetry, or the 
mechanisms to stabilize moduli. Hopefully further ingredients in the 
construction will help to improve the situation in the future.

\medskip

Our purpose in the present lecture is to show that there exist other 
setups in string theory, leading to constructions with features close to 
that of the Standard Model, and as satisfactory as (or even more than) the 
heterotic setup. In this lecture we will center on compactifications of 
type IIA theory with D6-branes wrapped on intersecting 3-cycles, 
extensively studied in the last two years 
\cite{bgkl,afiru,bkl,imr,rest,susy,bbkl,local}
\footnote{See \cite{bachas,orbif,magnetised} for earlier related work.}. 
The discussion is organized as follows: In Section \ref{branes} we briefly 
review D-branes in string theory, and we introduce intersecting D6-branes 
as a natural way to obtain charged 4d chiral fermions. In Section 
\ref{compact} we describe compactifications of type IIA theory with 
intersecting D6-branes, give rules to obtain their spectrum, and provide 
explicit examples for toroidal backgrounds. We briefly discuss the generic 
phenomenological features in this kind of construction, i.e. the natural 
predictions of the setup. In Section \ref{susy} we discuss the 
advantages/disadvantages of constructing supersymmetric vs 
non-supersymmetric models, and the different issues that arise in the 
latter case, and briefly describe some advanced models in each approach. 
Section \ref{conclu} contains some final comments.

\section{Intersecting D-branes}
\label{branes}

\subsection{D-branes in string theory}

New knowledge on string theory beyond perturbation theory has led to the 
introduction of new objects in string theory, D-branes, which provide a 
brand-new way of realizing non-abelian gauge symmetries in string theory. 
Therefore it is natural to explore the possibility of using them in 
searching for new setups with potential phenomenological application. We 
now review the main properties of D-branes from this biased viewpoint.

\medskip

\begin{figure}
\begin{center}
\centering
\epsfysize=3.5cm
\leavevmode
\epsfbox{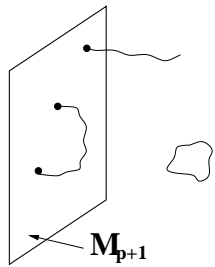}
\end{center}
\caption[]{\small Fluctuations of string theory around a D$p$-brane 
configuration are describe as open strings ending on its volume.}
\label{dp}
\end{figure}

Type II string theories contains certain `soliton-like' states in their 
spectrum, with $p+1$ extended dimensions, the $p$-branes. They were 
originally found as solutions of the low-energy supergravity equations of 
motion. Subsequently, it was realized \cite{polchinski} that certain of 
these objects (known as D$p$-branes) admit a fully stringy description, as 
$(p+1)$-dimensional subspaces on which open strings can end. Notice that 
these open strings are not present in the vacuum of the underlying string 
theory, but rather represent the fluctuations of the theory around the 
topological defect background. Namely, the closed string sector still 
describes the dynamics of the vacuum (gravitational interactions, etc), 
while open strings rather describe the dynamics of the object. The 
situation is shown in figure \ref{dp}

The spectrum of fluctuations of the theory in the presence of the 
D$p$-brane is obtained by quantizing closed strings and open strings 
ending on the D$p$-brane. Since the open string endpoints are fixed on 
the D-brane, the massless modes in the latter sector yield fields 
propagating on the $(p+1)$-dimensional D-brane world-volume $W_{p+1}$. 
For a single type II D$p$-brane in flat 10d space, such massless modes 
correspond to a $U(1)$ gauge boson, $9-p$ real scalars and some fermions. 
The scalars (resp. fermions) can be regarded as Goldstone bosons (resp. 
Goldstinos) of the translational symmetries (resp. supersymmetries) of the 
vacuum broken by the presence of the D-brane. The open string sector fills 
out a $U(1)$ vector multiplet with respect to the 16 supersymmetries 
unbroken by the D-brane.

\begin{figure}
\begin{center}
\centering
\epsfysize=3.5cm
\leavevmode
\epsfbox{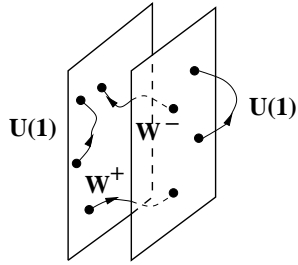}
\end{center}
\caption[]{\small Non-abelian gauge bosons in a configuration of 
coincident D-branes.}
\label{nonabelian}
\end{figure}

D$p$-branes are charged under the corresponding RR $(p+1)$-form $C_{p+1}$ 
of type  II string theory, via the minimal coupling $\int_{W_{p+1}}\, 
C_{p+1}$. Since flat D$p$-branes in flat space preserve 1/2 of the 32 
supercharges of the type II vacuum, such D-branes are BPS states, and 
their RR charge is related to their tension. This implies that there is no 
net force among parallel branes (roughly, gravitational attraction cancels 
against `Coulomb' repulsion due to their RR charge), so they can be 
superposed. The open string spectrum in a configuration of $n$ coincident 
D$p$-branes consists of $n^2$ sectors, corresponding to the $n\times n$ 
possible ways of choosing the D-brane on which the string starts and 
ends. This multiplicity renders interactions between open strings 
non-abelian, and the complete massless open string spectrum is given by 
$U(n)$ gauge bosons, $9-p$ adjoint scalars and adjoint fermions, filling 
out a $U(n)$ vector multiplet with respect to the 16 unbroken 
supersymmetries. The structure of gauge bosons for $n=3$ is shown in 
figure \ref{nonabelian}. As announced, D-branes provide a brand-new 
realization of non-abelian gauge symmetry in string theory. Indeed, the 
low-energy effective action for the massless open string modes is 
Dirac-Born-Infeld action, which at low energies (lowest derivative order) 
reduces to the Yang-Mills action. 

\subsection{Brane-worlds}

\begin{figure}
\begin{center}
\centering
\epsfysize=3.5cm
\leavevmode
\epsfbox{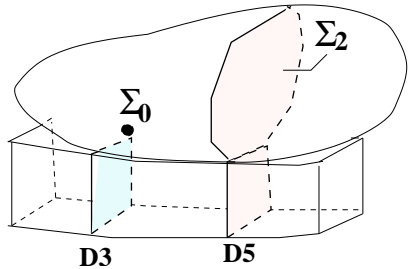}
\end{center}
\caption[]{\small Picture of D-brane worlds.}
\label{braneworld}
\end{figure}

Hence D-branes allow for gauge sectors localized on subspaces of spacetime, 
in a consistent and microscopically well-defined framework. This is a 
string theory realization of the brane-world idea \cite{aadd}. Namely we 
have a configuration where gravity propagates in 10d (with 4d gravity 
eventually reproduced at low energies via standard compactification on 
some internal space$\IX_6$); on the other hand, gauge interactions 
(hopefully, in a elaborated enough construction, with some gauge sector 
similar to the Standard Model) propagate just in the volume of a 
lower-dimensional subpace of spacetime. Starting with $(p+1)$-dimensional 
gauge interactions on a D$p$-brane, 4d gauge interactions can be obtained 
by considering the D$p$-brane volume to be of the form $\IM_4\times 
\Sigma_{p-3}$, with $\Sigma_{p-3}$ a $(p-3)$ closed subspace i.e. a 
$(p-3)$-cycle, of $\IX_6$. This is shown in picture \ref{braneworld}.

Since gauge and gravitational interactions arise in different string  
sectors (open vs closed) and propagate on different dimensions, the 
analysis of couplings and scales differs from that in heterotic. Before 
compactification, gravitational and gauge interactions are described by an 
effective action
\beqa
\int \, d^{10}x \, \frac{M_s^{\, 8}}{g_s^2} \, R_{10d} \, +
\int \, d^{p+1}x \, \frac{M_s^{\, p-3}}{g_s} \, F_{(p+1)d}^{\, 2}
\eeqa
where the powers of $g_s$ follow from the Euler characteristic of the
worldsheet which produces interactions for gravitons (sphere) and for 
gauge bosons (disk).
 
Upon compactification, the 4d action picks us volume factors and reads
\beqa
\int \, d^{4}x \, \frac{M_s^{\, 8}V_{6}}{g_s^2} \, R_{4d} \, +
\int \, d^{4}x \, \frac{M_s^{\, p-3} V_{\Sigma}}{g_s} \, F_{4d}^{\, 2}
\eeqa
This allows to read off the 4d Planck mass and gauge coupling, which are
experimentally measured.
\beqa
M_P^2 & = & \frac{M_s^{\, 8}V_{X_6}}{g_s^2} \simeq 10^{19} \, {\rm GeV}
\nonumber\\
1/g_{YM}^2 & = & \frac{M_s^{\, p-3} V_{\Sigma}}{g_s} \simeq 0.1
\eeqa
If the geometry is factorizable, we can split $V_{X_6}=V_{\Sigma}V_{\perp}$, 
with $V_{\perp}$ the transverse volume, and obtain
\beqa
M_{P}^2\, g_{YM}^2 \, = \, \frac{M_s^{11-p}V_{\perp}}{g_s}
\eeqa
This shows that it is possible to generate a large Planck mass in 4d with
a low string scale, by simply increasing the volume transverse to the
brane. In particular, it has been proposed to lower the string scale down 
to the TeV scale to avoid a hierarchy with the weak scale. The hierarchy 
problem is recast in geometric terms, namely the stabilization of the 
compactification size in very large volumes (the extreme case being around 
one millimiter for two large dimensions). Notice however that a low string 
scale is not compulsory in models with some solution to the hierarchy 
problem, e.g. supersymmetric models.

\subsection{D-branes and chirality}

\begin{figure}
\begin{center}
\centering
\epsfysize=3.5cm
\leavevmode
\epsfbox{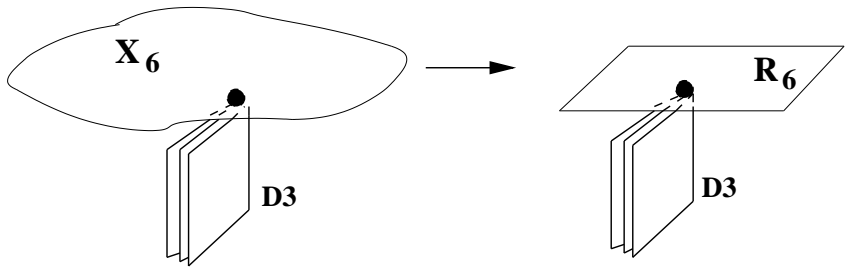}
\end{center}
\caption[]{\small Isolated D-branes at a smooth point in transverse space 
feel a locally trivial geometry and lead to non-chiral open string 
spectra.}
\label{zoom}
\end{figure}

We have remarked that D-branes in flat space preserve a lot of 
supersymmetry, namely 16 supercharges, which amounts to $\NN=4$ in 4d. 
This is too much to allow for a chiral open string spectrum. Indeed it is 
quite non-trivial to construct configurations of D-branes with chiral open 
string sector. One can heuristically show that {\em isolated} D-branes 
with a {\em smooth} transverse space automatically lead to non-chiral 
spectra \footnote{We are oversimplifying, since it may be possible to 
obtain chiral 4d fermions by turning on topologically non-trivial 
world-volume gauge backgrounds on $\Sigma_{p-3}$. We will not consider 
this possibility in the present discussion.}, as follows (see figure 
\ref{zoom}): The massless open string sector in a D-brane configuration 
in only sensitive to the local geometry around the D-branes. If they are 
isolated and sitting at a smooth point, the local structure is that of 
D-branes in flat space, which lead to non-chiral spectra.

There are two well-studied kinds of D-brane configurations which lead to 
chiral open string spectra, obtained by relaxing either of the above 
emphasized keywords.

{\bf a)} D-branes sitting at {\em singular} (rather than smooth) points 
in transverse space can lead to chiral open string spectra. The 
prototypical example is given by stacks of D3-branes sitting at 
the singular point of orbifolds of flat space, e.g. orbifold singularities 
$\IC^3/\IZ_N$, as studied in \cite{dm}. Such configurations have been used 
for phenomenological model building e.g. in \cite{singu}.

{\bf b)} Sets of intersecting D-branes (which are hence not isolated) can 
also lead to chiral fermions in the sector of open strings stretched 
between different kinds of D-brane \cite{bdl}. The chiral fermions are 
localized at the intersection of the brane volumes, in order to minimize 
its stretching. In the present lecture we will center on this kind of 
configuration.

\subsection{Intersecting D-branes}

\begin{figure}
\begin{center}
\centering
\epsfysize=3.5cm
\leavevmode
\epsfbox{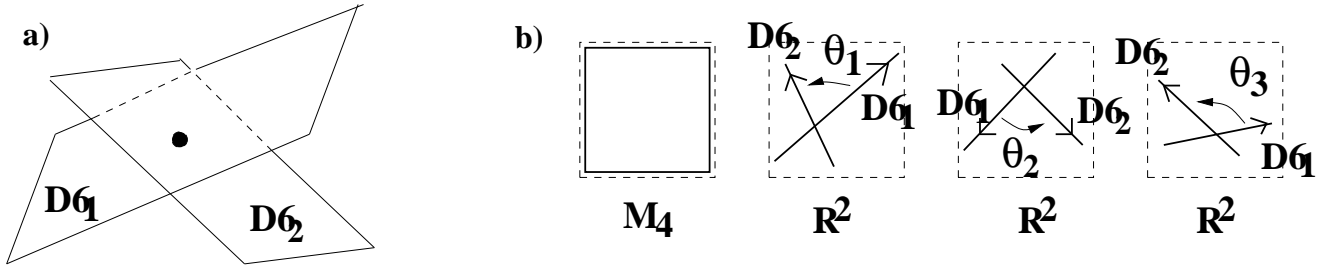}
\end{center}
\caption[]{\small Two picture of D6-branes intersecting over a 4d subspace 
of their volumes.}
\label{intersection}
\end{figure}

The basic configuration of intersecting D-branes leading to chiral 4d 
fermions at their intersection is two stacks of D6-branes in flat 10d 
intersecting over a 4d subspace of their volumes. Figures 
\ref{intersection} a, b provide  two pictorial representations of the 
configurations.

Chirality of the sector of open string stretching between the two 
D6-branes is consistent with the fact that any continuous motion of the 
branes (preserving the gauge symmetry) maintains the existence of an 
intersection; this corresponds to the fact that chiral particles at the 
intersection do not become massive upon deforming their effective action 
in a continuous fashion.

\medskip

The open string spectrum in a configuration of two stacks of ($n_1$ resp. 
$n_2$ coincident) D6-branes in flat 10d intersecting over a 4d subspace 
of their volumes consists of three open string sectors:

{\bf 6$_1$6$_1$} Strings stretching between D6$_1$-branes provide $U(n_1)$ 
gauge bosons, three real adjoint scalars and fermion superpartners, 
propagating over the 7d world-volume  of the D6$_1$-branes.

{\bf 6$_2$6$_2$} Similarly, strings stretching between D6$_2$-branes 
provide $U(n_2)$ gauge bosons, three real adjoint scalars and fermion 
superpartners, propagating over the D6$_2$-brane 7d world-volume.

{\bf 6$_1$6$_2$ + 6$_2$6$_1$} Strings stretching between both kinds of 
D6-brane lead to a 4d chiral fermion, transforming in the representation 
$(n_1,{\ov n}_2)$ of $U(n_1)\times U(n_2)$, and localized at the 
intersection \footnote{The chirality of the fermion is encoded in the 
orientation defined by the intersection; this will be implicitly taken 
into account in our discussion.}. 

In addition, the latter sector leads to scalar fields in the same 
bi-fundamental representation, and whose masses depend on the local 
geometry of the intersection. The lightest complex scalars have masses
\beqa
\alpha' M^2/2  & = & \frac 12 (\theta_1+\theta_2-\theta_3) \quad ; \quad
\frac 12 (\theta_1-\theta_2+\theta_3) \nonumber \\
&& \frac 12 (-\theta_1+\theta_2+\theta_3) \quad ; \quad 1-\frac 
12(\theta_1+\theta_2+\theta_3) 
\label{scmass}
\eeqa
where the $\theta_i$ are $1/\pi$ times the angles defined in Fig 
\ref{intersection}b, and taken between $-1$ and $1$.

In the generic case, there is no supersymmetry invariant under the two 
stacks of branes, and the open string sector at the intersection is 
non-supersymmetric. However, if $\theta_1\pm \theta_2 \pm \theta_3=0$ for 
some choice of signs, one of the scalars becomes massless, reflecting 
that the configuration is $\NN=1$ supersymmetric. $\NN=2$ supersymmetry 
arises if e.g. $\theta_3=0$ and $\theta_1\pm\theta_2=0$, while $\NN=4$ 
arises only for parallel stacks $\theta_i=0$.

\section{Four-dimensional models}
\label{compact}

Once we have succeeded in describing configurations of D-branes leading to 
charged chiral fermions, in this section we employ them in building  
models with 4d gravity and gauge interactions. Although intersecting  
D6-branes provide 4d chiral fermions already in flat 10d space, gauge  
interactions remain 7d and gravity interactions remain 10d unless we  
consider compactification of spacetime.

\begin{figure}
\begin{center}
\centering
\epsfysize=3.5cm
\leavevmode
\epsfbox{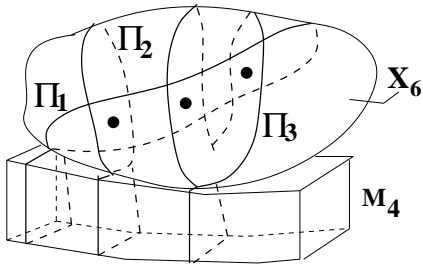}
\end{center}
\caption[]{\small Compactification with intersecting D6-branes wrapped on 
3-cycles.}
\label{compact}
\end{figure}     

The general kind of configurations we are to consider (see figure 
\ref{compact}) is type IIA string theory on a spacetime of the form $\IM_4
\times\IX_6$ with compact $\IX_6$, and with stacks of $N_a$ D6$_a$-branes 
with volumes of the form $M_4\times \Pi_a$, with $\Pi_a\subset \IX_6$ a 
3-cycles. It is important to realize that generically 3-cycles in a 6d 
compact space intersect at points, so the corresponding wrapped D6-branes 
will intersect at $M_4$ subspaces of their volumes. Hence, compactification 
reduces the 10d and 7d gravitational and gauge interactions to 4d, and 
intersections lead to charged 4d chiral fermions. Also, generically two 
3-cycles in a 6d space intersect several times, therefore leading to a 
replicated sector of opens strings at intersections. This is a natural 
mechanism to explain/reproduce the appearance of replicated families of 
chiral fermions in Nature!

\subsection{Toroidal models}

\begin{figure}
\begin{center}
\centering
\epsfysize=3.5cm
\leavevmode
\epsfbox{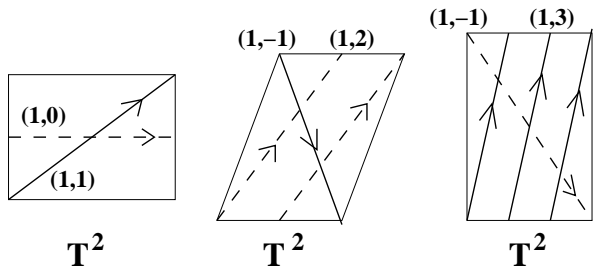}
\end{center}
\caption[]{\small Examples of intersecting 3-cycles in $\IT^6$.}
\label{wrapping}
\end{figure}

In this section we mainly follow \cite{afiru}, see also \cite{bgkl}.
To keep the configurations simple, we consider $\IX_6$ to be a six-torus 
factorized as $\IT^6=\IT^2\times \IT^2\times \IT^2$. Also for simplicity 
we take the 3-cycles $\Pi_a$ to be given by a factorized product of 
1-cycles in each of the 2-tori. For a 3-cycle $\Pi_a$, the 1-cycle in the 
$i^{th}$ 2-torus will be labeled by the numbers $(n_a^i,m_a^i)$ it wraps 
along the horizontal and vertical directions, see figure \ref{wrapping} 
for examples.

The intersection number is given by the product of the number of 
intersections in each 2-torus, and reads
\beqa
I_{ab}\, = \, (n_a^1 m_b^1-m_a^1n_b^1)\times (n_a^2 m_b^2-m_a^2 n_b^2)
\times (n_a^3 m_b^3-m_a^3 n_b^3)
\label{iab}
\eeqa

It is useful to introduce the 3-homology class $[\Pi_a]$ of the 3-cycle 
$\Pi_a$, which can be thought of as a vector of RR charges of the 
corresponding D6-brane. The 1-homology class of an $(n,m)$ 1-cycle in a 
2-torus is $n[a]+m[b]$, with $[a],[b]$ the basic homology cycles in 
$\IT^2$. For a 3-cycles with wrapping numbers $(n_a^i,m_a^i)$ we have
\beqa
[\Pi_a]\, =\, \otimes_{i=1}^3 \, (\, n_a^i\, [a_i]\, +\,m_a^i\,[b_i]\,)
\eeqa
The intersection number (\ref{iab}) is intersection number in homology, 
denoted $I_{ab}=[\Pi_a]\cdot[\Pi_b]$. This is easily shown using 
$[a_i]\cdot [b_j]=\delta_{ij}$ and linearity and antisymmetry of the
intersection pairing.

\subsubsection{Spectrum}

With the basic data defining the configuration, namely ${\bf N_a}$ 
D6$_a$-branes wrapped on 3-cycles ${\bf [\Pi_a]}$, with wrapping numbers 
${\bf(n_a^i,m_a^i)}$ on each $\IT^2$ and intersection numbers 
${\bf I_{ab}}$, we can compute the spectrum of the model. 

The closed string sector produces 4d $\NN=8$ supergravity. There exist 
different open string sectors:

{\bf 6$_a$6$_a$} String stretched among D6-branes in the $a^{th}$ stack 
produce 4d $U(N_a)$ gauge bosons, 6 real adjoint scalars and 4 adjoint 
Majorana fermions, filling out a vector multiplet of the 4d $\NN=4$ 
supersymmetry preserved by the corresponding brane.

{\bf 6$_a$6$_b$} + {\bf 6$_b$6$_a$} Strings stretched between the $a^{th}$ 
and $b^{th}$ stack lead to $I_{ab}$ replicated chiral left-handed 
\footnote{Negative intersection numbers lead to a positive number of 
chiral fermions with right-handed chirality.} fermions in the bifundamental 
representation $(N_a,{\ov N}_b)$. Additional light scalars may be present, 
with masses (\ref{scmass}) in terms of angles determined by the wrapping 
numbers and the $\IT^2$ moduli.

\medskip

Generalization for compact spaces more general than the 6-torus will be 
discussed in section \ref{beyond}.
We have therefore obtained a large class of four-dimensional theories with 
interesting non-abelian gauge symmetries and replicated charged chiral 
fermions. Hence compactifications with intersecting D6-branes provide a 
natural setup in which string theory can produce gauge sectors with the 
same rough features of the Standard Model. It is interesting to explore 
them further as possible phenomenological models, and  construct explicit 
examples with spectrum as close as possible to the Standard Model.

\begin{figure}
\begin{center}
\centering
\epsfysize=3cm
\leavevmode
\epsfbox{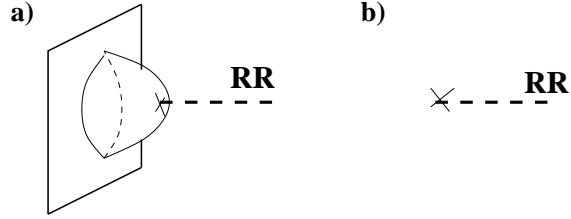}
\end{center}
\caption[]{\small Disk diagrams couple D-branes to closed strings, 
and lead to tadpoles for RR fields.}
\label{tadpole}
\end{figure}

\subsubsection{Cancellation of RR tadpoles}
\label{tadpoles}

\begin{figure}
\begin{center}
\centering
\epsfysize=3cm
\leavevmode
\epsfbox{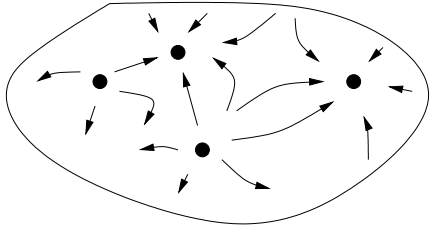}
\end{center}
\caption[]{\small In a compact space, fluxlines cannot escape and the 
total charge must vanish.}
\label{tadtwo}
\end{figure}

String theories with open string sectors must satisfy a crucial 
consistency condition, known as cancellation of RR tadpoles. As mentioned 
above, D-branes act as sources for RR $p$-forms via the disk coupling 
$\int_{W_{p+1}} C_p$, see fig \ref{tadpole}a. The consistency condition 
amounts to requiring the total RR charge of D-branes to vanish, as 
implied by Gauss law in a compact space (since RR field fluxlines cannot 
escape, figure \ref{tadtwo}). In our setup, the 3-cycle homology classes 
are vectors of RR charges, hence the condition reads
\beqa
[\Pi_{\rm tot}]\, =\, \sum_a \, N_a \, [\Pi_a]\, = \, 0
\label{chargezero}
\eeqa

Equivalently, the condition of RR tadpole cancellation can be expressed
as the requirement of consistency of the equations of motion for RR 
fields. In our situation, the terms of the spacetime action depending on 
the RR 7-form $C_7$ are
\beqa
S_{C7} & = & \int_{\IM_4\times \IX_6} H_8 \wedge * H_8 \, +\, \sum_a\,
N_a\, \int_{\IM_4\times \Pi_a} C_7 \, = \nonumber \\
&= & \int_{\IM_4\times \IX_6} C_7\wedge dH_2 \, +\, \sum_a \, N_a\, 
\int_{\IM_4\times \IX_6} C_7 \wedge \delta(\Pi_a)
\eeqa
where $H_8$ is the 8-form field strength, $H_2$ its Hodge dual, and 
$\delta(\Pi_a)$ is a bump 3-form localized on $\Pi_a$ in $\IX_6$. The 
equations of motion read
\beqa
dH_2\, = \, \sum_a\, N_a\, \delta(\Pi_a)
\eeqa
The integrability condition is obtained by taking this equation in 
homology, yielding (\ref{chargezero}).

It is useful to describe the latter interpretation of RR cancellation 
condition from the viewpoint of the compactified theory. In this language, 
there exist components of the KK reduction of $C_7$ leading to  4d RR 
4-forms. Taking a basis of 3-cycles in $\IX_6$, $[\Sigma_i]$, these are 
the zero modes $C_{4,i}=\int_{[\Sigma_i]} C_7$. Being 4d 4-forms, these 
fields do not have kinetic terms in their 4d action, and they only appear 
in linear tadpole terms of the form
\beqa
S_{C4,i}\, = \, [\Pi_a]\cdot [\Sigma_i]\, \int_{\IM_4} C_{4,i}
\eeqa
The equations of motion then imply that the coefficient of the tadpole 
must vanish $[\Pi_a]\cdot [\Sigma_i]=0$, namely it is not a condition on 
the field but a consistency condition for the model. For a complete basis 
of $[\Sigma_i]$, this implies (\ref{chargezero}).

\subsubsection{Anomaly cancellation}

Cancellation of RR tadpoles in the underlying string theory configuration 
implies cancellation of four-dimensional chiral anomalies in the effective 
field theory in our configurations. Recall that the chiral piece of the 
spectrum is given by $I_{ab}$ chiral fermions in the representation 
$(N_a,{\ov N}_b)$ of the gauge group $\prod_a U(N_a)$.

{\bf Cubic non-abelian anomalies}

The $SU(N_a)^3$ cubic anomaly is proportional to the number of 
fundamental minus antifundamental representations of $SU(N_a)$, hence it is 
proportional to $\sum_b I_{ab}N_b$. 

It is easy to check this vanishes due to RR tadpole cancellation: Starting 
with (\ref{chargezero}), we consider the intersection of $[\Pi_{\rm 
tot}]$ with any $[\Pi]$ to get
\beqa
0\, = \, [\Pi_a]\cdot \sum_b\,  N_b\, [\Pi_b]\, =\, \sum_b N_b I_{ab}
\eeqa
as claimed.

It is interesting to notice that RR tadpole cancellation is slightly 
stronger than cancellation of cubic non-abelian anomalies. In fact, the 
former requires that the number of fundamental minus antifundamentals 
vanishes even for the cases $N_a=1,2$, where no gauge theory anomaly 
exists. This observation will turn out relevant in phenomenological model 
building.

\medskip

\begin{figure}
\begin{center}
\centering
\epsfysize=3cm
\leavevmode
\epsfbox{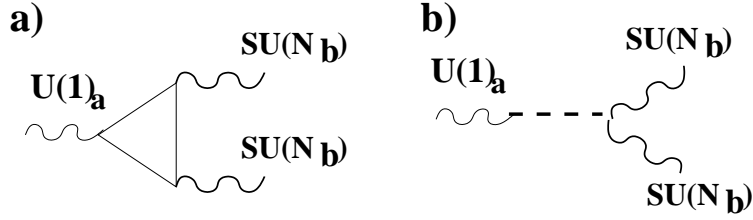}
\end{center}
\caption[]{\small Triangle and Green-Schwarz diagrams contributing to the 
mixed $U(1)$ - non-abelian anomalies.}
\label{gs}
\end{figure}

{\bf Cancellation of mixed anomalies}

The $U(1)_a$-$SU(N_b)^2$ mixed anomalies also cancel as a consequence of 
RR tadpole cancellation. They do so in a trickier way, namely the anomaly 
receives two non-zero contributions which cancel each other 
\footnote{Mixed gravitational triangle anomalies cancel automatically, 
without Green-Schwarz contributions.}, see fig \ref{gs}.

The familiar field theory triangle diagrams give a contribution which 
(after using RR tadpole conditions) is
\beqa
A_{ab}\, \simeq\,  N_a\, I_{ab}
\eeqa

On the other hand, the theory contains contributions from Green-Schwarz 
diagrams, where the gauge boson of $U(1)_a$ mixes with a 2-form which 
subsequently couples to two gauge bosons of $SU(N_b)$.

The coupling between $U(1)_a$ and a 2-form $\int_{[\Pi_a]} C_5$ arises 
from the coupling
\beqa
N_a\, \int_{D6_a}\, C_5\wedge \tr F_a\, \stackrel{KK}{\longrightarrow}
\,N_a\, \int_{\IM_4}\, B_a\wedge \tr F_a\, 
\eeqa
There are also scalars $\phi_b=\int_{[\Pi_b]} C_3$ which couple to 
$SU(N_b)$ gauge bosons via
\beqa
\int_{D6_b}\, C_3\wedge \tr F_a^{\, 2}\, \stackrel{KK}{\longrightarrow}
\, \int_{\IM_4}\, \phi_b \tr F_b^{\, 2} 
\eeqa
One can get the duality relation $dB_a\, = \,I_{ab} \, * d\phi_b$ and 
check that the contribution to the anomaly is proportional to
\beqa
A'_{ab}\, \simeq\,  -N_a\, I_{ab}
\eeqa
leading to a cancellation between both kinds of contributions.

\begin{figure}
\begin{center}
\centering
\epsfysize=1.3cm
\leavevmode
\epsfbox{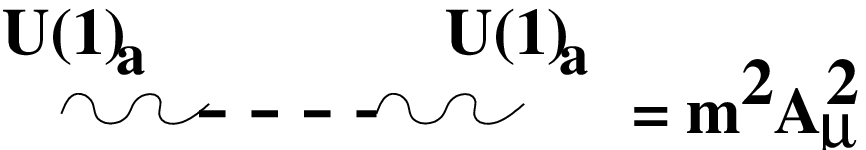}
\end{center}
\caption[]{\small The $B\wedge F$ couplings lead to a $U(1)$ gauge boson 
mass term.}
\label{mass}
\end{figure}

An important observation is that any $U(1)$ gauge boson with $B\wedge F$ 
couplings gets massive, with mass roughly of the order of the string 
scale, see fig \ref{mass}. Such $U(1)$'s disappear as gauge symmetries 
from the low-energy effective field theory, but remain as global 
symmetries, unbroken in perturbation theory.

\subsubsection{A Standard Model - like example}

Consider a configuration of D6-branes on $\IT^6$ defined by the following 
wrapping numbers

\begin{center}
\begin{tabular}{cccc}
$N_1=3$ & $(1,2)$ & $(1,-1)$ & (1,-2) \\
$N_2=2$ & $(1,1)$ & $(1,-2)$ & (-1,5) \\
$N_3=1$ & $(1,1)$ & $(1,0)$ & (-1,5) \\
$N_4=1$ & $(1,2)$ & $(-1,1)$ & (1,1) \\
$N_5=1$ & $(1,2)$ & $(-1,1)$ & (2,-7) \\
$N_6=1$ & $(1,1)$ & $(3,-4)$ & (1,-5) 
\end{tabular}
\end{center}

The intersection numbers are

\begin{center}
\begin{tabular}{cccccc}
$I_{12}=3$ & $I_{13}=-3$ & $I_{14}=0$ & $I_{15}=0$ & $I_{16}=-3$ \\
$I_{23}=0$ & $I_{24}=6$ & $I_{25}=3$ & $I_{26}=0$ & $I_{34}=-6$ \\
$I_{35}=-3$ & $I_{36}=0$ & $I_{45}=0$ & $I_{46}=6$ & $I_{56}=3$ 
\end{tabular}
\end{center}

A $U(1)$ linear combination, playing the role of hypercharge, remains 
massless
\beqa
Q_Y\,  =\, -\frac 13 \, Q_1 \, - \, \frac 12 \, Q_2\, - \, Q_3\, -\, Q_5
\eeqa
The chiral fermion spectrum, with charges with respect to the Standard 
Model - like gauge group, is
\beqa
& SU(3)\times SU(2)\times U(1)_Y \times \ldots & \nonumber\\
& 3 (3,2)_{1/6} + 3({\ov 3},1)_{-2/3} + 3({\ov 3},1)_{1/3} + 6(1,2)_{-1/2} 
+ & \nonumber \\
& + 3(1,2)_{1/2} + 6(1,1)_1 + 3(1,1)_{-1} + 9(1,1)_0 &
\eeqa
It is a quite nice toy model, very close to the Standard Model spectrum, 
but with additional matter, in particular six extra $SU(2)$ doublets.
The origin of the later can be understood from RR tadpole cancellation: 
As mentioned above, they require the number of fundamentals to equal that 
of antifundamentals, even for $SU(2)$. Since, to obtain three 
left-handed quarks we have chiral fermions in $3(3,{\ov 2})$, the full 
model must contain nine fields $(1,2)$, three of which correspond to left 
handed leptons and six of them remaining as exotic additional matter. In 
section \ref{imrmodels} we will see that more advanced models, including 
orientifold planes, can avoid this difficulty.

\subsection{Generalization beyond torus}
\label{beyond}

Clearly the above setup is not restricted to toroidal compactifications. 
Indeed one may take take any compact 6-manifold as internal space, for 
instance a Calabi-Yau threefold, which would lead to 4d $\NN=2$ 
supersymmetry in the closed string sector.
In this situation we should pick a set of 3-cycles $\Pi_a$ on which we 
wrap $N_a$ D6-branes (for instance special lagrangian 3-cycles of $\IX_6$ 
if we are interested in preserving supersymmetry), making sure they 
satisfy the RR tadpole cancellation condition $\sum_a N_a [\Pi_a]=0$.

The final open string spectrum (for instance, in the case of 
supersymmetric wrapped D6-branes) arises in two kinds of sectors

{\bf 6$_a$-6$_a$} Leads to $U(N_a)$ gauge bosons ($\NN=1$ vector 
multiplets in the supersymmetric case) and $b_1(\Pi_a)$ real adjoint 
scalars (chiral multiplets in susy case).

{\bf 6$_a$-6$_b$+6$_b$-6$_a$} We obtain $I_{ab}$ chiral fermions in the 
representation $(N_a,{\ov N}_b)$ (plus light scalars, massless in 
supersymmetry preserving intersections). Here 
$I_{ab}=[\Pi_a]\cdot[\Pi_b]$.

Notice that the chiral spectrum is obtained in terms of purely topological 
information of the configuration, as should be the case. 

\medskip

The phenomenology of non-toroidal models will be quite similar to that of 
toroidal compactifications with D-branes, see next subsection. Thus, the 
later are in any event good toy model for many features of general 
compactifications with intersecting branes. This is particularly 
interesting since it is relatively difficult to construct explicit 
configurations of intersecting D6-branes in Calabi-Yau models (although 
some explicit examples have been discussed in \cite{bbkl,local}).

\subsection{Phenomenological features}

We now turn to a brief discussion of the phenomenological properties 
natural in this setup \cite{afiru}.

$\bullet$ As we soon discuss, most models constructed in the literature 
are non-supersymmmetric. It is possible but notoriously more difficult to 
construct fully $\NN=1$ supersymmetric models. Therefore, unless 
alternative solutions to the hierarchy model are provided, the best 
proposal is to consider non-susy models to have a low string scale 
$M_s\simeq$ TeV to avoid hierarchy.

$\bullet$ The proton is stable in these models, since the $U(1)$ within 
the $U(3)$ color factor plays the role of baryon number, and is preserved 
as a global symmetry, exactly unbroken in perturbation theory. 
Non-perturbative effects breaking it arise from euclidean D2-branes 
wrapped on 3-cycles, and have the interpretation of spacetime gauge theory 
instantons, hence reproducing the non-perturbative breaking of baryon 
number in the Standard Model. This feature of proton stability is most 
welcome in models with low fundamental scale.

$\bullet$ These models do {\em not} have a natural gauge coupling 
unification, even at the string scale. Each gauge factor has a gauge 
coupling controlled by the volume of the wrapped 3-cycle
\beqa
\frac{1}{g_{YM,a}^{\;2}}\, =\, \frac{M_S^3\, V_{\Pi_a}}{g_S}
\eeqa
Gauge couplings are related to geometric volumes, hence their experimental 
values can be adjusted/reproduced in concrete models, rather than 
predicted by the general setup.

\begin{figure}
\begin{center}
\centering
\epsfysize=3.5cm
\leavevmode
\epsfbox{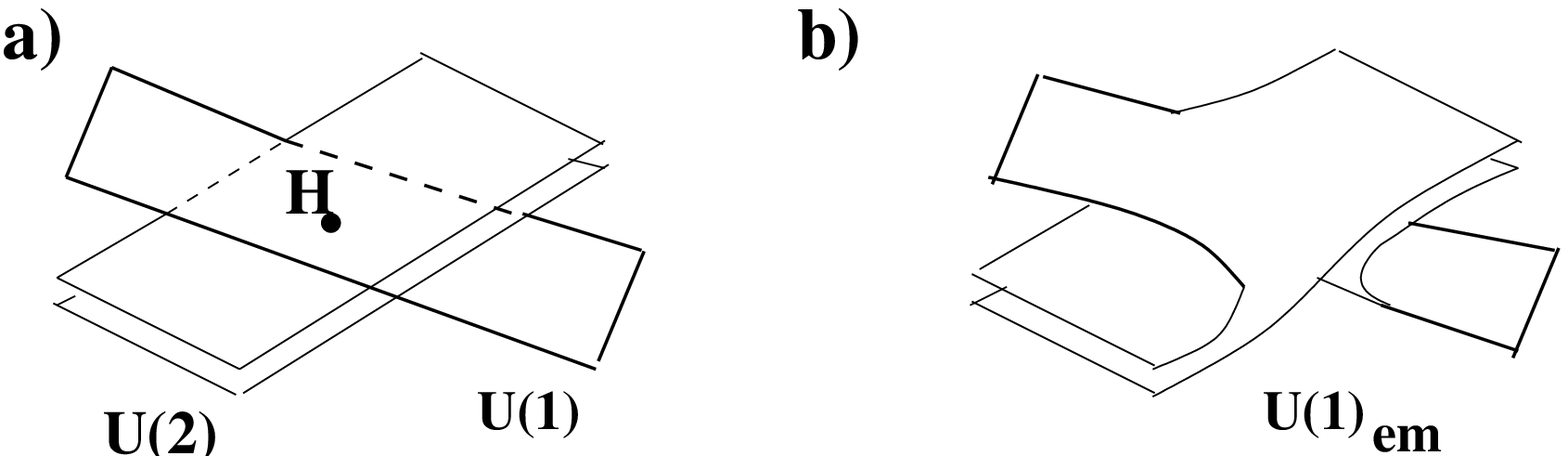}
\end{center}
\caption[]{\small 
.}
\label{higgs}
\end{figure}     

$\bullet$ There exists a geometric interpretation for the spontaneous 
electroweak symmetry breaking. In explicit models, the Higgs scalar 
multiplet arises from the light scalars at intersections (which are 
massless in susy cases, and massive or tachyonic in non-susy cases). In 
intersecting D-brane configurations, vevs for scalars at intersections 
parametrize the possibility of recombining two intersecting cycles into a 
single smooth one, as shown in figure \ref{higgs}. In the process, the 
gauge symmetry is reduced, corresponding to a Higgs mechanism in the 
effective field theory. See \cite{cimhiggs} for further discussion.

\begin{figure}
\begin{center}
\centering
\epsfysize=3.5cm
\leavevmode
\epsfbox{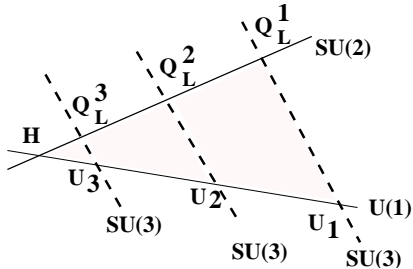}
\end{center}
\caption[]{\small Geometric origin of the hierarchy of Yukawa couplings 
for different generations.}
\label{yukawa}
\end{figure}

$\bullet$ There is a natural exponential hierarchy of the Yukawa 
couplings. Yukawa couplings among the scalar Higgs and chiral fermions at 
intersections arise at tree level in the string coupling from open string 
worldsheet instantons; namely from string worldsheets spanning the 
triangle with vertices at the intersections and sides on the D-branes. 
Their value is roughly given by $e^{-A}$, with $A$ the triangle area in 
string units \footnote{This is a good approximation only for large 
triangles, otherwise further worldsheet instanton contributions as well as 
fluctuations should be considered.}.
Since different families are located at different intersections, their 
triangles have areas increasing linearly with the family index, 
leading to an exponential Yukawa hierarchy, see fig \ref{yukawa}. See 
\cite{yuk} for further analysis of yukawa couplings in explicit models.

\section{Supersymmetry}
\label{susy}

In this section we discuss the supersymmetry or not of models of 
intersecting D6-branes, and the role of supersymmetry in phenomenological 
model building with these configurations.

The first observation we can make is that all models above constructed are 
non-supersymmetric. One simple way to see it is that we start with type 
IIA string theory compactified on $\IX_6$, and introduce D6-branes. Since 
RR tadpole cancellation requires that the total RR charge vanishes, we are 
forced to introduce objects with opposite RR charges, in a sense branes 
and antibranes, a notoriously non-supersymmetric combination.

An equivalent derivation of the result is as follows: If we would succeed 
in constructing a supersymmetric configuration of D6-branes, the system as 
a whole would be a supersymmetric BPS state of type IIA on $\IX_6$. Since 
for a BPS state the tension is proportional to the RR charge, and the 
latter vanishes due to RR tadpole cancellation, the tension of the state 
must vanish. The only D6-brane configuration with zero tension is having 
no D6-brane at all. Hence the only supersymmetric configuration would be 
just type IIA on $\IX_6$, with no brane at all.

These arguments suggest a way out of the impasse. In order to obtain 
$\NN=1$ supersymmetric compactifications we need to introduce objects with 
negative tension and negative RR charge, and which preserve the same 
supersymmetry as the D6-branes. Such objects exist in string theory and 
are orientifold 6-planes, O6-planes. Introduction of these objects leads 
to an interesting extension of the configurations above constructed, and 
will be studied in section \ref{oplanes}. In particular we will use them 
to construct supersymmetric compactifications with intersecting D6-branes.

\subsection{Issues on non-supersymmetric models}

Before doing that, it is interesting to consider diverse difficulties one 
encounters in non-supersymmetric models, and the extent to which they can 
be satisfactorily solved.

\subsubsection{Tachyons at intersections}

Even if each stack of D6-branes preserve some supersymmetry, the 
preserved susy may be different for different stacks. In these situations, 
the spectrum of open strings at intersections is generically 
non-supersymmetric, and some light scalar at the intersection may be 
tachyonic. Such tachyons signal an instability against recombination of 
the two sets of intersecting branes into a single smooth one. The 
initially considered configuration is not really stable.

There are two possible ways to face these tachyons. The first possibility 
is to consider configurations where all scalar fields at all intersections
are non-tachyonic, leading to non-supersymmetric models which are 
nevertheless stable against small perturbations. These are at best local 
minima, since as argued above, there always exists a global 
supersymmetric minimum, which is the type IIA vacuum, and corresponds to 
having no D6-brane at all. The metastable vacua may tunnel (by nucleation 
of D8-brane bubbles, see footnote 8 in \cite{coninst}) to this global 
minimum; they however provide reasonable enough models if sufficiently 
long-lived.

A second possibility is to employ tachyons of this kind, with the right 
quantum numbers, to trigger electroweak symmetry breaking. Although the 
string scale should be higher than the weak scale, the gauge symmetry 
breaking scale can be adjusted to around 100 GeV by fine-tuning the angles 
at the corresponding intersection. See \cite{cimhiggs} for further 
discussion of the Higgs as a tachyon.

\subsubsection{Hierarchy problem}

Despite their simplicity for model building, non-supersymmetric models of 
intersecting D6-branes on $\IT^6$ suffer a hierarchy problem. In order to 
avoid large corrections to the weak scale, it is natural to lower the 
string scale to around 1 TeV. The large value of the 4d Planck scale 
should therefore arise as a derived value due to some large internal 
dimension. However, if one attempts to increase any direction in $\IT^6$, 
too light KK replicas of Standard Model gauge bosons are generated, 
because the corresponding D6-branes span all such directions. In other 
words, there is no direction in $\IT^6$ which is transverse to all 
Standard Model D6-branes.

This is however a mild problem, in the sense that it is very specific to 
toroidal models. It is easy to imagine more general Calabi-Yau 
compactifications where the 3-cycles wrapped by the Standard Model branes 
are localized in a small region of the Calabi-Yau, and may be of 
small volume even for large Calabi-Yau volume. Some steps towards explicit 
realizations of this idea have been taken in \cite{bbkl,local}.

\subsubsection{NSNS tadpoles}

The most difficult issue in non-supersymmetric string vacua (even 
in general situations, not restricted to the case of intersecting D6-brane 
models) is the existence of uncancelled tadpoles for NSNS fields 
(graviton, dilaton, moduli). 

In contrast with RR tadpoles, NSNS tadpoles do {\em not} signal an 
inconsistency of the theory. Reviewing the arguments in section 
\ref{tadpoles}, they differ in the fact that NSNS fields do have kinetic 
terms, hence the equations of motion can be satisfied by balancing the 
tadpole term with the kinetic term. The existence of a nonzero tadpole in 
a particular background means that it is not a correct background for the 
theory, and should be consequently corrected. The corrected backgrounds 
are usually curved geometries which must be computed in an $\alpha'$ 
expansion, resulting in quite complicated final models, sometimes (but 
not always \cite{dmt}) involving naked singularities and strong coupling 
regions (see \cite{nsns}).

In view of this situation two different attitudes have been taken. It is 
fair to say that none of them solves the issue of NSNS tadpoles in a fully 
satisfactory manner; after all, it is related to the difficult and 
unsolved questions of breaking of supersymmetry.

$\bullet$ Stick to $\NN=1$ supersymmetric models, to avoid NSNS tadpoles 
in the string construction. This approach obviously `ignores' 
supersymmetry breaking in the real world, since once this is implemented 
the issue of NSNS tadpoles would re-arise.

$\bullet$ Ignore the issue of NSNS tadpoles completely! This approach is 
justified by recognizing that we really do not know the complete equations 
of motion of string theory, hence it is possible that a background which 
does not solve the classical equations of motion still provides a good 
approximation to the solution to the full string equations of motion.

A good field theory analogy has been proposed by C. Bachas \cite{bachas}. 
Consider a 4d $U(1)$ gauge field theory with charged fermions and a 
charged massless complex scalar with potential $V=|\phi|^4$. The 
solution to the classical equations of motion is $\phi=0$, around which 
the theory has unbroken gauge symmetry and massless fermion and scalar 
excitations. A configuration not solving the equations of motion is 
$\phi=v\neq 0$, which sits on the slope of the scalar potential and 
therefore leads to a non-zero $\phi$-tadpole. Ignoring the tadpole, the 
physics around this configuration is that of spontaneously broken gauge 
symmetry and massive fermions and scalar. 

Interestingly enough, in many situations the second configuration provides 
the best approximation to the real physics of the system once quantum 
corrections are taken into account, since they may develop a mass term for 
$\phi$, and a minimum of the corrected scalar potential with non-zero vev 
for $\phi$. The complete physics of the system is better approximated by 
a configuration {\em not} solving the classical equations of motion.
Analogously \cite{bachas}, {\em `Perhaps the price for getting a good 
description of the low energy world from string theory may be to allow for 
small metric, dilaton and moduli tadpoles in the classical description of 
the groundstate'}. 

\medskip

Our purpose in the remaining is to discuss some of the most interesting 
models constructed within each approach. Before doing that, it will be 
convenient to introduce some background material on certain useful 
objects, the O6-planes.

\subsection{Detour: Orientifold 6-planes}
\label{oplanes}

Consider  type IIA theory on $\IX_6$, and mod out the configuration by 
$\Omega$ (worldsheet orientation reversal) times a $\IZ_2$ symmetry $g$ of 
$\IX_6$. The set of fixed points of $g$ form an orientifold plane, namely 
a subspace of spacetime where the orientation of the string can flip.

A simple example, on which we center henceforth, is provided by 
$\IX_6=\IT^6$ with $g$ given by 
the action $y_i\to -y_i$, where $y_i$ are the vertical direction on each 
$\IT^2$. This is a symmetry for rectangular two-tori (see figure 
\ref{osix}), or for two-tori tilted by a specific angle \cite{bkl}, see 
below. 
For figure \ref{osix}, the set of fixed points is given by $x_i$ 
arbitrary, $y_i=0,R_{y_i}/2$, hence has 8 components. Since each has seven 
dimensions (counting also the $\IM_4$ piece), they correspond to 
O6-planes wrapped on the 3-cycle with wrapping numbers $(n_i,m_i)=(1,0)$.

\begin{figure}
\begin{center}
\centering
\epsfysize=3.5cm
\leavevmode
\epsfbox{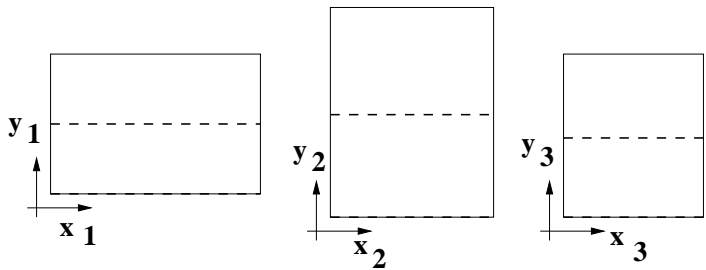}
\end{center}
\caption[]{\small Orientifold 6-planes in the orientifold quotient of 
IIA on $\IT^6$ by $\Omega g$, with $g:y_i\to -y_i$.}
\label{osix}
\end{figure}     

\begin{figure}
\begin{center}
\centering
\epsfysize=2.5cm
\leavevmode
\epsfbox{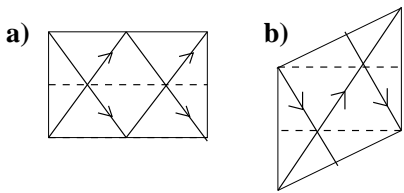}
\end{center}
\caption[]{\small Cycles and their orientifold images in a rectangular and 
tilted 2-tori.}
\label{osixtwo}
\end{figure}

One may consider introducing D6-branes in the above orientifold quotient 
of IIA on $\IT^6$. The configurations are described as in the toroidal 
situation, by specifying the multiplicities $N_a$ and wrapping numbers 
$(n_a^i,m_a^i)$ of the D6-brane stacks, with two novelties: 

First, in order to have a configuration invariant under $\Omega g$ we 
need to introduce orientifold images (denoted D6$_{a'}$-branes) of the 
D6$_a$-branes. They have multiplicity $N_a$ and wrapping numbers 
$(n_a^i,-m_a^i)$ for rectangular 2-tori, or $(n_a^i,-n_a^i-m_a^i)$ for 
tilted tori, see fig \ref{osixtwo}. To simplify the latter case we 
introduce ${\tilde m}_a=m_a+n_a/2$, so that branes and images have 
wrapping numbers $(n_a,{\tilde m}_a)$ and $(n_a,-{\tilde m}_a)$ respectively.

Second, the O6-planes preserve the same supersymmetry as D6-branes on the 
same 3-cycle, and they carry negative tension and negative RR charge ($-4$ 
in D6-brane charge units). Consequently, they contribute to the RR 
tadpole condition, which now reads (for rectangular tori, i.e. 8 
O6-planes)
\beqa
\sum_a\, N_a\, [\Pi_a] \, + \, \sum_a\, N_a\, [\Pi_{a'}] \, - \,4\times 
8\, 
[\Pi_{O6}]\, = \, 0
\eeqa

\medskip

The open string spectrum in orientifolded models now contains new sectors. 
It is given by

{\bf aa+a'a'} Contains $U(N_a)$ gauge bosons and superpartners

{\bf ab+ba+b'a'+a'b'} Contains $I_{ab}$ chiral fermions in the 
representation $(N_a,{\ov N}_b)$, plus light scalars.

{\bf ab'+b'a+ba'+a'b} Contains $I_{ab'}$ chiral fermions in the 
representation $(N_a,N_b)$, plus light scalars.

{\bf aa'+a'a} Contains certain numbers $n_{\Ysymm}$, $n_{\Yasymm}$ of 
chiral fermions in the two-index symmetric resp. antisymmetric
representations, with $n_{\Ysymm}+n_{\Yasymm}=I_{aa'}$.

As expected, the new RR tadpole conditions in the presence of O6-planes  
guarantee the cancellation of 4d anomalies of the new chiral spectrum, in 
analogy with the toroidal case \footnote{In the orientifold case, mixed 
gravitational anomalies may receive Green-Schwarz contributions 
\cite{susy}}. 

\subsection{Advanced model building}

\subsubsection{Models in the non-susy setup}
\label{imrmodels}

In this section we consider some of the phenomenologically most 
interesting examples of non-supersymmetric models, `ignoring' the issue of 
the corresponding NSNS tadpoles. What follows is a brief sketch of the 
ideas in \cite{imr}, to which we refer the reader for details.

Consider type IIA on $\IT^6$ modded out by $\Omega g$ with $g:y_i\to 
-y_i$, and introduce four stacks of D6-branes (plus images) leading to a 
gauge group
\beqa
U(3)_a\times U(2)_b \times U(1)_c \times U(1)_d
\eeqa
and with the only non-zero intersection numbers being

\begin{center}
\begin{tabular}{cccc}
$I_{ab}=1$ & $I_{ab'}=2$ & $I_{ac}=-3$ & $I_{ac'}=-3$ \\
$I_{bd}=0$ & $I_{bd'}=-3$ & $I_{cd}=-3$ & $I_{cd'}=3$
\end{tabular}
\end{center}

One can check that the number of fundamentals equals that of  
antifundamentals even for $SU(2)$ and $U(1)$ factors. For instance, for 
$SU(2)$ we have left handed quarks in the representation $2(3,{\ov 2})$ + 
$({\ov 1},2)$ and leptons in $3(1,2)$. Notice that the model does not 
require six extra doubles, avoiding this feature of the purely 
toroidal model thanks to the existence of the new kind of bifundamental 
representation $(N_a,N_b)$.

In \cite{imr} a large class of explicit 3-cycles in $\IT^6$ with those 
intersection numbers were constructed, and for which there is a unique 
massless $U(1)$ linear combination which plays the role of hypercharge. 
The resulting complete chiral spectrum is
\beqa
& SU(3)_c\times SU(2)_w \times U(1)_Y & \nonumber \\
3 & \times  [\, (3,2)_{1/6}\, +\, ({\ov 3},1)_{1/3}\, + \, ({\ov 
3},1)_{-2/3} \, + & \nonumber \\
& +\, (1,2)_{-1/2} \, + \, (1,1)_{1}\, + \, (1,1)_0 \, ] &
\eeqa
Namely {\em just} the Standard Model (plus right handed neutrinos).

Much of the recent activity in string model building has centered on the 
search of similarly successful models in non-toroidal geometries 
\cite{bbkl,local}, or supersymmetric realizations of these intersection 
numbers. In many respects, the search is on!

\subsubsection{Supersymmetric models}

In this section we review the first supersymmetric 4d chiral model of 
intersecting D6-branes, in \cite{susy} to which we refer the reader for 
details (see \cite{susy2} for additional models).

In order to obtain supersymmetric models, one needs a sufficient number of 
O6-planes in the construction. One of the simplest possibilities is the 
$\Omega g$ orientifold of the $\IT^6/(\IZ_2\times \IZ_2)$ orbifold. The 
rules for model building are similar to the orientifolded tori case, with 
additional care to implement the $\IZ_2$ orbifold actions. 

Skipping the details, let us simply mention that the conditions to have 
all D6-brane intersections supersymmetric and cancel RR tadpoles are 
restrictive and do not allow for a systematic classification of models. 
Among the most interesting ones in \cite{susy}, let us simply mention a 
configuration of six D6-brane stacks (plus $\Omega g$ images), with 
wrapping numbers
\beqa
N_1=8 \quad & (0,1)\times (0,-1)\times (2,-1) \quad & \to U(1) \nonumber 
\\
N_2=2 \quad & (1,0)\times (1,0)\times (2,-1) \quad & \nonumber \\
N_3=4 \quad & (1,0)\times (1,-1)\times (1,1) \quad & \to U(2) \nonumber \\
N_4=2 \quad & (1,0)\times (0,1)\times (0,-1) \quad &  \nonumber \\
N_5=6+2 \quad & (1,-1)\times (1,0)\times (1,0) \quad & \to U(3)\times U(1) 
\nonumber \\
N_6=4 \quad & (0,1)\times (1,0)\times (0,-1) \quad &
\eeqa
The configuration is $\NN=1$ supersymmetric for a suitable choice 
of the $\IT^2$ radii ratios.  

As shown above, the model contains a subgroup with the non-abelian 
structure of the Standard Model. It also contains one linear 
combination of the above $U(1)$'s which remains massless, and which 
plays the role of hypercharge. The final chiral spectrum, with respect to
this Standard Model subgroup is
\beqa
& SU(3)_c\times SU(2)_w \times U(1)_Y & \nonumber \\
3 & \times  [\, (3,2)_{1/6}\, +\, ({\ov 3},1)_{1/3}\, + \, ({\ov 
3},1)_{-2/3} \, + & \nonumber \\
& +\, (1,2)_{-1/2} \, + \, (1,1)_{1}\, + \, (1,1)_0 \, ] &
\eeqa
plus {\em chiral} exotic matter. Chiral exotics render the model 
unrealistic, but the nice Standard Model - like subsector make it a good 
toy model of supersymmetric model building with intersecting D6-branes. 
Some of its phenomenological features have been described in 
\cite{upenn}.

\section{Final comments}
\label{conclu}

We have shown that compactifications with D6-branes wrapped on intersecting 
3-cycles provide a large class of 4d models with non-abelian gauge 
symmetries and chiral fermions. A first lesson to be drawn is that 
D-branes in string theory {\em do} allow for phenomenologically realistic 
consistent theories realizing the brane world idea, and therefore providing 
an alternative to other scenarios, like weakly coupled heterotic theory or 
Horava-Witten theory compactifications.

We have briefly discussed some of the interesting phenomenological 
features of the present setup, like the possibility to lower the string 
scale, the stability of the proton, and the interesting hierarchy of 
yukawa couplings. Also, although we did not discuss them, the models (in 
particular the supersymmetric oned) have a rich structure of duality 
relations with other interesting string and M theory constructions, like 
M-theory on $G_2$ manifolds, or D-branes with worldvolume magnetic fluxes.
Hopefully further research of these models and their properties will shed 
light on the difficult issues that remain to be addressed, most notably 
the issue of supersymmetry breaking and/or NSNS tadpoles.

\medskip

{\Large \bf Acknowledgements}

I would like to thank G. Aldazabal, M. Cvetic, S. Franco, L. E. 
Ib\'a\~nez, R. Rabad\'an and G. Shiu for collaboration in these topics, 
and the organizers of the TMR meeting for creating an stimulating 
environment. I also thank M. Gonz\'alez for her kind support and 
encouragement.

\end{document}

\begin{figure}
\begin{center}
\centering
\epsfysize=3.5cm
\leavevmode
\epsfbox{zzz.eps}
\end{center}
\caption[]{\small 
.}
\label{zzz}
\end{figure}